\documentclass[onecollarge,runningheads]{FPhys}       

\smartqed  
\usepackage{graphicx}
\usepackage{upgreek}
\usepackage{hyperref}
\DeclareGraphicsExtensions{.jpeg,.jpg,.ps,.eps,.pdf}

\newcommand{\bra}[1]{\langle{#1}|}
\newcommand{\ket}[1]{|{#1}\rangle}
\newcommand{\ip}[1]{\langle{#1}\rangle}

\newcommand{\beq}{\begin{equation}}
\newcommand{\eeq}{\end{equation}}
\newcommand{\beqa}{\begin{eqnarray}}
\newcommand{\eeqan}{\end{eqnarray*}}
\newcommand{\beqan}{\begin{eqnarray*}}
\newcommand{\eeqa}{\end{eqnarray}}
\newcommand{\non}{\nonumber}
\newcommand{\hs}[1]{\hspace{#1cm}}
\renewcommand{\theequation}{\arabic{equation}}
\newcommand{\eq}[1]{(\ref{#1})}

\newcommand{\ary}[2]{
                \begin{array}{#1} #2 \end{array}}
\newcommand{\mat}[2]{\left[
                \begin{array}{#1} #2 \end{array}\right]}

\newcommand{\Hf}{H_{F}}
\newcommand{\Hb}{H_{B}}
\newcommand{\Uf}{U_{F}}
\newcommand{\Ub}{U_{B}}

\newcommand{\bK}{{\overline K}}

\newcommand{\sfc}[2]{\mbox{$\frac{#1}{#2}$}}

\journalname{\small{Found Phys (2011)}}
\DOI{\small \href{http://dx.doi.org/10.1007/s10701-011-9568-x}{DOI: 10.1007/s10701-011-9568-x}}

\begin{document}

\title{T violation and the unidirectionality of time}

\titlerunning{Found Phys (2011)}        

\author{Joan A. Vaccaro\vspace{25mm} 
}

\authorrunning{Found Phys (2011)} 

\institute{Joan A. Vaccaro \at
           Centre for Quantum Dynamics, Griffith
           University, 170 Kessels Road, Brisbane 4111, Australia.\\
           \email{J.A.Vaccaro@griffith.edu.au}
}

\date{Received: 22 September 2010 / Accepted: 8 May 2011}

\maketitle

\large
\begin{abstract} {An increasing number of experiments at the Belle, BNL,
CERN, DA$\Upphi$NE and SLAC accelerators are confirming the violation of time
reversal invariance (T). The violation signifies a fundamental asymmetry
between the past and future and calls for a major shift in the way we think
about time. Here we show that processes which violate T symmetry induce
destructive interference between different paths that the universe can take
through time. The interference eliminates all paths except for two that
represent continuously forwards and continuously backwards time evolution.
Evidence from the accelerator experiments indicates which path the universe
is effectively following. This work may provide fresh insight into the
long-standing problem of modeling the dynamics of T violation processes.  It
suggests that T violation has previously unknown, large-scale physical
effects and that these effects underlie the origin of the unidirectionality
of time. It may have implications for the Wheeler-DeWitt equation of
canonical quantum gravity. Finally it provides a view of the quantum nature
of time itself.\\}

\keywords{CP violation \and T violation \and Kaons \and Arrow of time \and
Quantum interference \and Quantum foundations \and  Wheeler-DeWitt equation}
\end{abstract}

\section{Introduction}

The physical nature of time has been an enigma for centuries. The main tools
for discussing its unidirectionality are the phenomenological arrows of time.
A great deal of progress has been made in recent decades in linking various
arrows together \cite{Price}.  A notable exception is the matter-antimatter
arrow which arises from the violation of charge-parity conjugation invariance
(CP) in meson decay \cite{Aharony1,Aharony2,Berger}. CP violation, which was
first discovered by Cronin, Fitch and coworkers \cite{Cronin} in 1964 in the
decay of neutral kaons (K mesons), provides a clue to the origin of the
large-scale matter-antimatter imbalance of the universe \cite{Sakharov}. The
combined CPT invariance has been confirmed in kaon decay to a high degree of
precision \cite{Pavlopoulos} suggesting that neutral kaon decay also violates
T invariance. To date no violation of CPT has been observed in B meson decay
\cite{Lusiani} and so the CP violation in B meson decay also appears to be
consistent with T violation. But more importantly, experimental evidence has
confirmed the direct T violation in K meson decay \cite{Angelopoulos}
independently of CPT invariance. Nonetheless, T violating processes are
relatively rare and the magnitudes of the violations are relatively small,
and so despite their important asymmetric temporal nature, they are often
regarded as having little impact on the nature of time. As a result, the
physical significance of T violation has remained obscure.

The CP violation due to the weak interaction is described in the Standard
Model of particle physics by the Cabibbo-Kobayashi-Maskawa (CKM) matrix
\cite{Cabibbo,Kobayashi}. Nevertheless, a fundamental and intriguing question
remains open. In principle, the CKM model provides the basis for the time
evolution of mesons in terms of Schr\"{o}dinger's equation and an associated
Hamiltonian $H$. The evidence of T violation implies that $THT^{-1}\ne H$
where $T$ represents the time reversal operation \cite{Wigner}. But how
should one incorporate the \textit{two} Hamiltonians, $H$ for time evolution
towards our future and $THT^{-1}$ for evolution towards our past, in one
quantum equation of motion? At issue is the fact that Schr\"{o}dinger's
equation describes consistent unitary evolution in both directions of time:
the unitary evolution by negative time intervals merely ``back tracks''
exactly the same time evolution described by positive time intervals. Yet,
paradoxically, the evidence of T violation processes appears to suggests that
Nature does not work in this way in general.

Currently, to model T violating processes we must choose the direction of time evolution and a
specific version, say $H$, of the Hamiltonian in order to write down the Schr\"{o}dinger
equation. By applying the time reversal operation we then obtain the Schr\"{o}dinger equation
involving the $THT^{-1}$ version of the Hamiltonian for evolution in the opposite direction of
time. But at any point we are restricted to having a dynamical equation of motion for a
specific direction of time evolution and a corresponding specific version of the Hamiltonian.
We do not have a dynamical equation of motion for the situation where the {\it direction of
time evolution cannot be specified}.  In this case there is no argument for favoring one
version of the Hamiltonian over the other and so if one version of the Hamiltonian is to
appear in the equation of motion then so must the other. This problem becomes critical when we
attempt to describe the universe as a closed system as needed in cosmology, for then the
direction of time evolution cannot be specified because a closed system precludes any external
clock-like device for use as a reference for the direction of time. There is, therefore, no
satisfactory quantum formalism for describing a universe that exhibits T violation processes.
This failure of standard quantum theory is all the more serious given that CP and, by
implication, T violation play a critical role in baryogenesis in the early history of the
universe \cite{Sakharov,belle}. It is clearly an important problem that lies at the very
foundations of quantum theory. Its resolution calls for a major shift in the way we think
about time and dynamical equations of motion.

We address this issue by deriving a single quantum dynamical equation which incorporates the
two versions of a T-violating Hamiltonian. We begin by using Feynman's sum over histories
\cite{Feynman} method to construct the set of all possible paths that the universe can take
through time. The set represents every possible evolution zigzagging forwards and backwards
through time. We then show how T violation induces destructive interference that eliminates
most of the possible paths that the universe can follow. Effectively only two main paths,
representing continuously forwards and continuously backwards time evolution, survive the
interference. Physical evidence from accelerator experiments allows us to to {\it distinguish
between the two directions of time} and determine which path the universe is effectively
following.  In short, T violation is shown to have {\it a previously unknown large-scale
physical effect that underlies the unidirectionality of time}. This is the main result of this
work. We end with a discussion that includes the potential impact for cosmology and the
Wheeler-DeWitt equation in canonical quantum gravity.

\section{Possible paths through time}

Consider a system, which we shall refer to simply as the universe, whose
composition in terms of size, matter and fields is consistent with the
visible portion of the physical universe. For clarity we assume that the
universe is closed in the sense that it does not interact with any other
physical system and further, that there is no physical system external to the
universe. This means that there is no clock external to the universe and so
our analysis needs to be unbiased with respect to the direction of time
evolution. Nevertheless it is convenient to differentiate two directions of
time as ``forward'' and ``backward'' associated with evolution in the
positive$-t$ and negative$-t$ directions of the time axis, respectively,
although we stress that neither direction necessarily has any connection to
our actual everyday experience. Also, we begin our analysis with the universe
in a state $\ket{\psi_0}$ which we shall call the ``origin state'' without
reference to the direction of time. With this in mind we express the
evolution of the universe in the forwards direction over the time interval
$\tau$ as $\ket{\psi_{F}(\tau)}=U_{F}(\tau)\ket{\psi_{0}}$ where
\begin{equation}
    \label{defn UF}
    U_{F}(\tau)=\exp(-i\tau H_{F})
\end{equation}
is the forwards time evolution operator, and $H_{F}$ is the Hamiltonian for forwards time
evolution. Throughout this article we use units in which $\hbar=1$. If time reversal symmetry
were obeyed, the evolution in the opposite direction of time would be given by $\exp(i\tau
H_{F})$.  To accommodate the time asymmetry of T violation processes, we need to replace
$H_{F}$ in this expression with its time reversed form. Thus we define the state of the
universe after backwards evolution over a time interval of the same magnitude by
$\ket{\psi_{B}(\tau)}=U_{B}(\tau)\ket{\psi_{0}}$ where
\begin{equation}
    \label{defn UB}
    U_{B}(\tau)=\exp(i\tau H_{B})\ ,
\end{equation}
and $H_{B}=TH_{F}T^{-1}$ is the Hamiltonian for backwards time evolution.
This gives $U_{B}(\tau)=TU_{F}(\tau)T^{-1}$, in accord with Wigner's
definition of time reversal \cite{Wigner}.  In other words, $H_F$ and $H_B$
are the generators of time translations in the positive$-t$ and negative$-t$
directions, respectively.

The parameter $\tau$ in these expressions represents a time interval that
could be observed using clock devices which are internal to the universe. A
simple model of a clock for our purposes is a device which evolves unitarily
through a sequence of orthogonal states at sufficient regularity to give a
measure of a duration of time to any desired accuracy. The orthogonal states
would represent the eigenstates of a pointer observable which acts a
reference for time. A harmonic oscillator is a device of this kind
\cite{Barnett}. In order to avoid ambiguities in the definition of time
intervals for different time directions, we assume that the Hamiltonian
$H^{({\rm clk})}$ describing our clock model is T invariant and so
$T\,H^{({\rm clk})}\,T^{-1}=H^{({\rm clk})}$ during its normal operation.
This gives an unambiguous operational meaning of the parameter $\tau$ as a
time interval.  We shall refer to the readings of such clocks as ``clock
time''.

The matrix elements $\bra{\phi}U_{F}(\tau)\ket{\psi_{0}}$ and $\bra{\phi
}U_{B}(\tau)\ket{\psi_{0}}$ represent the probability amplitudes for the universe in state
$\ket{\psi_{0}}$ to evolve over the time interval $\tau$ to $\ket{\phi}$ via two paths in time
corresponding to the forwards and backwards directions, respectively. Given that we have no
basis for favoring one path over the other, we follow Feynman \cite{Feynman} and attribute an
equal statistical weighting to each. Thus, {\it the total probability amplitude for the
universe to evolve from one given state to another is proportional to the sum of the
probability amplitudes for all possible paths through time between the two states.} In the
current situation we have only two possible paths and so the total amplitude is proportional
to
\begin{equation}
   \label{add amplitudes}
   \bra{\phi}U_{F}(\tau)\ket{\psi_{0}}+\bra{\phi}U_{B}
   (\tau)\ket{\psi_{0}}=\bra{\phi}U_{F}(\tau)+U_{B}(\tau
  )\ket{\psi_{0}}\ .
\end{equation}
This result holds for all states $\ket{\phi}$ of the universe and so the unbiased time
evolution of $\ket{\psi_{0}}$ over the time interval $\tau$ can be written as
\begin{equation}
  \label{1 step bievolution}
  \ket{\Psi(\tau)}=\left[{U_{F}(\tau)+U_{B}
  (\tau)}\right]\ket{\psi_{0}}
\end{equation}
which we call the \textit{symmetric time evolution} of the universe. To reduce unnecessary
detail we use un-normalized vectors to represent states of the universe. We could have
included in \eq{add amplitudes} an additional phase factor $e^{i\theta}$ to represent an
arbitrary relative phase between the two paths, such as $\bra{\phi}U_{{
F}}(\tau)\ket{\psi_{0}}+e^{i\theta}\bra{\phi}U_{B}(\tau)\ket{\psi_{0}}$, but the factor has no
net physical effect for our analysis and so we omit it.

It follows that the symmetric time evolution of the universe in state $\ket{\Psi (\tau)}$ over
an additional time interval of $\tau$ is given by
\beqan
    \ket{\Psi (2\tau)}&=&\left[ {U_{F} (\tau)+U_{B}
   (\tau)} \right]\ket{\Psi (\tau)}\\
    &=&\left[ {U_{F} (\tau)+U_{B} (\tau)} \right]^{2}\ket{\psi_{0}}\ .
\eeqan
Repeating this for $N$ such time intervals yields
\begin{equation}
    \label{N step biev = (U+U)^N}
   \ket{\Psi (N\tau)}=\left[U_{F} (\tau)+U_{B} (\tau)  \right]^{N}\ket{\psi _{0}}\ .
\end{equation}

\begin{figure}
\begin{center}
  \includegraphics[width=100mm]{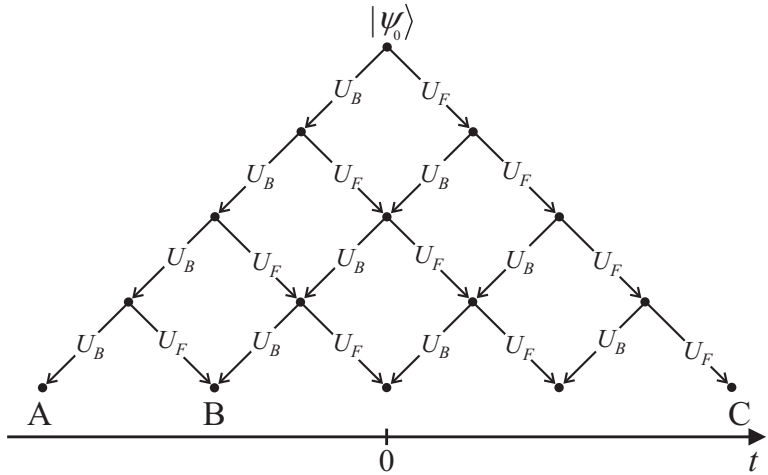}
\end{center}
\caption{\label{fig:tree with no interference}Binary tree representation of
the generation of the state  $\ket{\Psi(N\tau)}$ from the origin state
$\ket{\psi_0}$ according to \eq{N step biev = (U+U)^N} with $N=4$. States are
represented as nodes (solid discs) and unitary evolution by links (arrows)
between them. The root node (at the top) represents the origin state
$\ket{\psi_0}$ and the leaf nodes (on the bottom row) represent components of
the state $\ket{\Psi(4\tau)}$ where, for example, the node labeled A
represents the state $S_{4,0}\ket{\psi_0}=U_{B}^4\ket{\psi_0}$, B represents
the state
$S_{3,1}\ket{\psi_0}=[U_{B}^3U_{F}+U_{B}^2U_{F}U_{B}+U_{B}U_{F}U_{B}^2+U_{F}U_{B}^3]\ket{\psi_0}$
and C represents $S_{0,4}\ket{\psi_0}=U_{F}^4\ket{\psi_0}$. The time axis
labeled $t$ at the bottom of the figure represents clock time which increases
from left to right. Clock time is measured by clock devices whose Hamiltonian
$H^{\rm (clk)}$ is T invariant.}
\end{figure}

Figure \ref{fig:tree with no interference} gives a graphical interpretation of this
result in terms of a binary tree. It is useful to write the expansion of the product on
the right side as
\begin{equation}
    \label{N step biev = S_N-n,n}
    \ket{\Psi (N\tau)}=\sum\limits_{n=0}^N {S_{N-n,n}} \;\ket{\psi
    _{0}}
\end{equation}
where $S_{m,n} $ represents a sum containing $({_{\ \ n}^{n+m}})$ different terms each
comprising $n$ factors of $U_{F} (\tau)\;$ and $m$ factors of $U_{{B}} (\tau)$, and where
$({{}_j^k})=k!/[(k-j)!j!]$ is the binomial coefficient. $S_{m,n} $ is defined by the
recursive relation
\begin{equation}
    \label{defn S_m,n}
    S_{m,n} =\sum\limits_{k=0}^m {S_{m-k,n-1} U_{F} (\tau)U_{B}
    (k\tau)}
\end{equation}
with $S_{m,0}=U_{B}(m\tau)$ and gives, for example, $S_{m,1}=\sum_{k=0}^m {U_{{
B}}[(m-k)\tau ]U_{F}(\tau)U_{B}(k\tau)}$.

The expression $\bra{\phi}S_{N-n,n}\ket{\psi_{0}}$ represents the evolution of the universe
from $\ket{\psi_{0}}$ to $\ket{\phi}$ over a set of $({_n^N}\!)$ paths through time, where
each path comprises a total of $n$ steps in the forwards direction and $N-n$ steps in the
backwards direction. The set of paths includes all possible orderings of the forwards and
backwards steps. This set of paths is the focus of the remaining analysis.

First consider the size of the time steps $\tau$. To explore the consequences of the limit
$\tau\to 0$ of infinitely-small time steps for a fixed total time $t_{\rm tot}$, we set
$\tau=t_{\rm tot}/N$ and consider \eq{N step biev = (U+U)^N} for increasing values of $N$.  We
can write $2^{-N}[U_{F} (\tau)+U_{B}(\tau)]^N=\{\exp[-i{1\over 2}(H_{ F}-H_{B})\tau]+{\cal
O}(\tau^2)\}^N$ which becomes $\exp[-i{1\over 2}(H_{F}-H_{ B})t_{\rm tot}]$ as $N\to\infty$.
In this limit, the universe evolves according to the Hamiltonian ${1\over 2}(H_{F}-H_{B})$.
This means that isolated subsystems of the universe that obey T invariance, such as our model
of a clock, would not exhibit any evolution. As clocks are taken to measure time intervals,
the universe as a whole would not exhibit dynamics in the conventional sense in this $\tau \to
0$ limit. The lack of dynamics is an unphysical result if our system is to model the visible
universe. This suggests that the time interval $\tau$ should be a small non-zero number for a
universe-like system. Accordingly, in the following we set the value of $\tau $ to be the
smallest physically-reasonable time interval, the Planck time, i.e. $\tau \approx 5\times
10^{-44}{s}$.

We now derive a more manageable expression for $S_{m,n} $. The first step is to reorder
the expression so that all the $U_{B} $ factors are to the left of the $U_{{F}} $ factors
using the Zassenhaus formula \cite{Suzuki}. This yields
\beq
   \label{S_m,n=U_BU_F sum commutator}
   S_{m,n} = U_{B} (m\tau)U_{F} (n\tau)\sum\limits_{v=0}^m
   \cdots \sum\limits_{\ell =0}^s \sum\limits_{k=0}^\ell{\exp \left[
   {(v+\cdots +\ell +k)\tau ^{2}[{H}_{F} ,{H}_{B} ]} \right]\exp
   [{\cal O}(\tau ^{3})]}
\eeq
where there are $n$ summations on the right side and $[A,B]$ is the
commutator of $A$ and $B$. \ref{app-reordering} gives the details of the
calculation. Next, the analysis is made simpler by expressing the summand in
the eigenbasis of the Hermitian operator $i[{H}_{F} ,{H}_{B} ]$. For this we
use the resolution of the identity given by ${{\bf 1}}=\int {\overline
{\Pi}(\lambda)d\lambda} $ where the measure $d\lambda $ accommodates both
continuous and discrete spectra and $\overline {\Pi
}(\lambda)$ is the projection operator that projects onto the state space spanned by the
eigenstates of $i[{H}_{F} ,{H}_{B} ]$ with eigenvalue $\lambda $. The degeneracy of the
eigenvalue $\lambda $ is given by the density function $\rho (\lambda )={\rm
Tr}[\overline {\Pi}(\lambda)]$ where ${\rm Tr}$ is the trace. To emphasize the degeneracy
of $\lambda $ we write the identity operator as
\begin{equation}
   \label{resolution of id}
   {{\bf 1}}=\int {\rho (\lambda)\Pi (\lambda)d\lambda}
\end{equation}
where $\Pi(\lambda)=\overline {\Pi}(\lambda)/\rho (\lambda)$ is an operator with unit
trace. Thus, for example, $i[{H}_{F} ,{H}_{B} ]$ $=\int{\lambda \rho (\lambda )\Pi
(\lambda)d\lambda} $. Multiplying \eq{S_m,n=U_BU_F sum commutator} on the right by
\eq{resolution of id} and ignoring the term of order $\tau ^{3}$ gives
\begin{equation}
   \label{S_m,n=U_BU_F int I(lambda)}
   S_{m,n} =U_{B} (m\tau)U_{F} (n\tau)\int {I_{m,n} (\lambda
  )\rho (\lambda)\Pi (\lambda)d\lambda}
\end{equation}
where
\begin{equation}
    \label{I(lambda)=big sum}
    I_{m,n} (\lambda)=\;\sum\limits_{v=0}^m {\cdots \sum\limits_{\ell =0}^s
    {\sum\limits_{k=0}^\ell {\exp \left[ {-i\,(v+\cdots +\ell +k)\tau
    ^{2}\lambda} \right]}}} \quad .
\end{equation}
In the following analysis terms of order $\tau ^{3}$ are negligible and are
ignored. Performing the algebraic manipulations described in
\ref{app-simplifying} eventually yields
\begin{equation}
   \label{I(lambda)=simple}
   I_{m,n} (\lambda)=\frac{\prod\nolimits_{q=0}^{n-1} {\left\{ {\,1-\exp [-i(n+m-q)\tau
   ^{2}\lambda ]} \right\}}}{\prod\nolimits_{q=1}^n {\left[ {1-\exp (-iq\tau ^{2}\lambda)}
   \right]}}\ .
\end{equation}

There are three important points to be made about the results in Eqs.~(\ref{S_m,n=U_BU_F int
I(lambda)}) and (\ref{I(lambda)=simple}). The first is that if the universe's origin state
$\ket{\psi _{0}}$ is a $\lambda =0$ eigenstate of the operator $i\,[{H}_{F} ,\,{H}_{ B} ]$
then the integral in \eq{S_m,n=U_BU_F int I(lambda)} can effectively be replaced with $I_{m,n}
(0)\rho (0)\Pi (0)$ and the resulting evolution is equivalent to the T invariant case. In
order to focus on the implications of T violation we assume that the universe's origin state
$\ket{\psi _{0}}$ has a zero overlap with all $\lambda =0$ eigenstates, i.e.
\begin{equation}
   \label{Pi(0)|psi_0>=0}
   \Pi (0)\ket{\psi _{0}}=0\ .
\end{equation}
We call this the {\it nonzero eigenvalue} condition for convenience.  We look at the
implications of relaxing this condition in Section \ref{sec: relax condt nonzero
eigenvalues}.  The second point is a useful symmetry property of $I_{m,n} (\lambda)$.
Multiplying both the numerator and denominator of the right side of \eq{I(lambda)=simple}
by $\prod\nolimits_{k=1}^{n+m} {[1-\exp (-ik\tau^{2}\lambda)]} $, canceling like terms
and judiciously relabeling indices yields the symmetry
\begin{equation}
    \label{symmetry I_m,n=I_n,m}
    I_{m,n} (\lambda)=I_{n,m} (\lambda)\quad .
\end{equation}
For the third point, it is straightforward to show for $n\le m$ that the zeroes of the
denominator of $I_{m,n} (\lambda)$ in \eq{I(lambda)=simple} do not result in
singularities, because if any factor $[1-\exp (-iq\tau ^{2}\lambda)]$ in the denominator
is zero for $\lambda =\lambda _{0} $, there is a corresponding factor $\{1-\exp
[-i(n+m-\ell)\tau ^{2}\lambda ]\}$ in the numerator, where $\ell =n+m-jq$ and $j$ is a
positive integer, for which
\[
\frac{1-\exp [-i(n+m-\ell)\tau ^{2}\lambda ]}{1-\exp (-iq\tau ^{2}\lambda
)}\to 1
\]
as $\lambda \to \lambda_{0}$. The symmetry of $I_{m,n} (\lambda)$ in \eq{symmetry
I_m,n=I_n,m} means that the same result holds for the complementary case $n>m$.

\section{Interference between paths}

We now analyze the behavior of $S_{N-n,n} $ in \eq{N step biev = S_N-n,n} in
some detail. Let $t_{F} =n\tau $ and $t_{B} =(N-n)\tau $ be the aggregate
times that the universe evolves forwards and backwards, respectively, where
$N$ is the total number of steps and $t_{\rm tot} =t_{F} +t_{B} =N\tau $ is
the corresponding total time.

\begin{figure}
\begin{minipage}{45mm}
\vspace{-3.5cm}\caption{\label{fig:Integrand function I}Comparison of the
function $|I_{N-n,n}(\lambda)|$ (solid curves) with the quadratic
approximation (dashed curves) for various values of $N$ and $n$. All curves
have been scaled to give a peak value of 1.}
\end{minipage}
\hspace{5mm}
\begin{minipage}{100mm}
  \includegraphics[width=100mm]{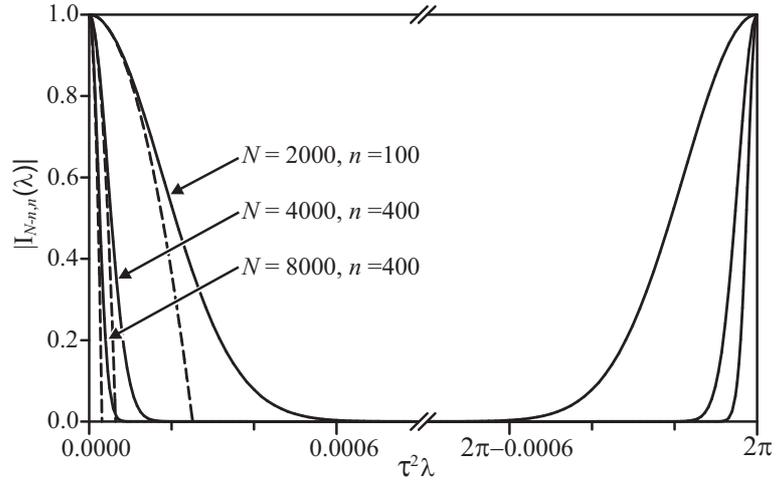}
\end{minipage}
\end{figure}

Certain features of $I_{N-n,n} (\lambda)$ are more apparent when its modulus $\vert
I_{N-n,n} (\lambda)\vert $ is written as the product of diffraction grating interference
functions, i.e.
\[
    \left| {I_{N-n,n} (\lambda)} \right|=\left| {\prod\nolimits_{q=1}^n
    {\frac{\sin (\alpha _{q} \beta _{q} \tau ^{2}\lambda)}{\sin (\beta _{q}
    \tau ^{2}\lambda)}}} \right|\ ,
\]
where $\alpha_{q} =\frac{1 }{ q}(N+1-q)$ and $\beta _{q} =\frac{q }{ 2}$.
$|I_{N-n,n} (\lambda)|$ is a periodic function with a period of $2\pi /\tau
^{2}$ and has maxima at $\lambda=k 2\pi/\tau^{2}$ of $\Pi _{q=1}^n \alpha
_{q} =({_{n}^N})$ for integer $k$. For the times $t_{F} =n\tau $ and $t_{B}
=(N-n)\tau $ to be physically measurable intervals, both $n$ and $(N-n)$ need
be very much larger than unity. In these cases we can treat
$|I_{N-n,n}(\lambda)|$ as comprising narrow peaks at $\lambda =k2\pi /\tau
^{2}$, for integer $k$, and being negligible elsewhere. Expanding
$|I_{N-n,n}(\lambda)|$ as a power series in $\lambda $ gives the quadratic
approximation $|I_{m,n}(\lambda)|\approx (^N_n)[1-\frac{1}{
{24}}n(N-n)(N+1)\tau ^{4}\lambda^{2}]$. Fig.~\ref{fig:Integrand function I}
illustrates the accuracy of the approximation near $\lambda =0$. The zeroes
of the quadratic approximation can be used as a reasonable estimate of the
width $W_{N-n,n} $ of the peaks. This estimate is given by
\[
W_{N-n,n} =\sqrt {\frac{24}{\tau ^{4}n(N-n)(N+1)]}} \quad .
\]

To estimate typical values for the eigenvalues $\lambda$ we take neutral kaon
evolution as a prototypical T-violating process. We first consider the
problem for a single kaon for which we label the eigenvalues and Hamiltonians
with a superscript $(1)$. The eigenvalues $\lambda^{(1)}$ of
$i[{H}_{F}^{(1)},{H}_{B}^{(1)}]$ in the 2 dimensional subspace spanned by the
kaon and anti-kaon states can be found using the phenomenological model of
Lee and Wolfenstein \cite{Lee}. In \ref{app-eigenvalues} we show that this
gives $\lambda ^{(1)}\approx \pm 10^{17}\,{s}^{-2}$ (in units where
$\hbar=1$) using empirical values of Yao {\it et al.} \cite{Yao}.

Next, we estimate the eigenvalues of the operator $i[{H}_{F},{H}_{B}]$ for the universe
containing $M$ kaons. For this we use the fact that the Hamiltonian of a non-interacting set
of subsystems is simply the sum of the Hamiltonians of the subsystems. Assuming that the
collection of kaons in the universe do not interact we can approximate each eigenvalue
$\lambda $ of the operator $i\,[{H}_{F} ,\,{H}_{{B}}]$ as a sum $\lambda =\sum_j {\lambda
_j^{(1)}}$ where $\lambda_j^{(1)} $ is an eigenvalue of $i[{H}_{F}^{(1)},{H}_{ B}^{(1)}]$ for
the $j$th kaon. The values of $\lambda $ range from $-M|\lambda ^{(1)}|$ to $M|\lambda
^{(1)}|$. The degeneracy of the eigenvalue $\lambda=k|\lambda ^{(1)}|$ for integer $k$ is
given by the density function $\rho(\lambda)=M!/(\frac{1 }{ 2}M-k)!(\frac{1 }{ 2}M+k)!$. Note
that $2^{-M}\rho (\lambda)$ is approximately a Gaussian distribution over $\lambda $ with a
mean of zero and standard deviation of $\lambda_{ SD} =\frac{1 }{ 2}\sqrt M|\lambda ^{(1)}|$
and so we shall take the typical values of $\lambda $ to range over $-\lambda_{\rm SD} ,\ldots
,\lambda_{\rm SD}$. Setting $M$ to be a fraction, $f<1$, of the average total number of
particles in the universe, say $10^{80}$, gives $M\approx f10^{80}$ and so $\lambda_{{\rm
SD}}\approx \sqrt f 10^{57}\,{s}^{-2}$.

\begin{figure}
\begin{center}
  \includegraphics[width=100mm]{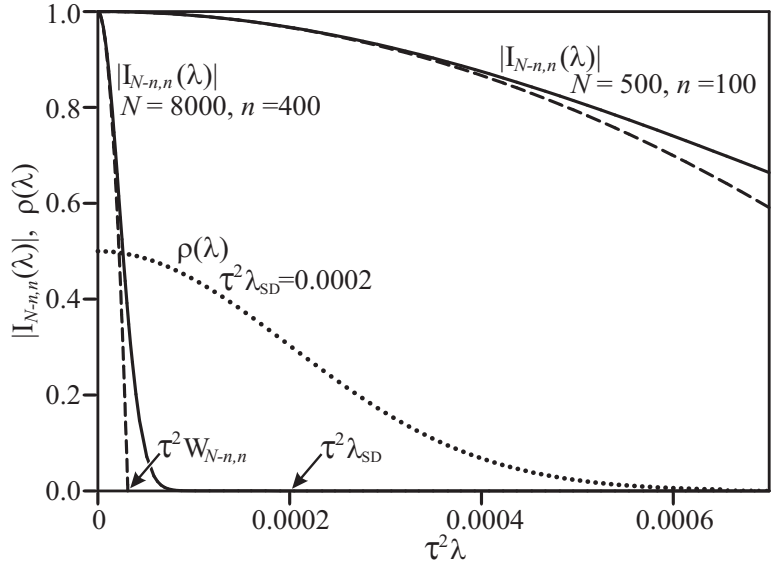}
\end{center}
\caption{\label{fig:interference}Comparison of the density function $\rho (\lambda)$ (dotted curve)
with the integrand function $|I_{N-n,n} (\lambda)|$ (solid curve) and its approximation (dashed curve). The curves representing
$|I_{N-n,n} (\lambda)|$ and its approximation have been
scaled to give a peak value of 1, whereas for clarity, the curve representing $\rho
(\lambda)$ has been scaled to give a peak value of 0.5. The standard deviation of the
density function is given by $\lambda _{\rm SD} =0.0002/\tau ^{2}$. The curve $|I_{N-n,n} (\lambda)|$ for $N=500$ and $n=100$ has a relatively
broad peak whereas for $N=8000$ and $n=400$ the peak is relatively narrow compared to the
peak in the density function curve.}
\end{figure}

There are two important limiting cases to consider that depend on the
relative sizes of the width parameters $W_{N-n,n} $ and $\lambda_{\rm SD} $
of $I_{N-n,n} (\lambda)$ and $\rho (\lambda)$, respectively, as illustrated
in Fig.~\ref{fig:interference}. The first is the regime given by $W_{N-n,n}
\gg\lambda_{\rm SD} $, i.e. relatively broad peaks of $I_{N-n,n} (\lambda )$.
In this limit $I_{N-n,n} (\lambda)$ can be treated as approximately constant
over the range $|\lambda|<\lambda_{\rm SD} $ and so the integral in
\eq{S_m,n=U_BU_F int I(lambda)} is well approximated by $(^N_n){{\bf 1}}$
where ${{\bf 1}}$ is the identity operator. This implies that there is
negligible destructive interference between the paths represented by
$\bra{\phi }S_{N-n,n} \ket{\psi_{0}}$. The second case is the regime
$W_{N-n,n} \ll \lambda_{\rm SD}$, i.e. relatively narrow peaks of
$I_{N-n,n}(\lambda)$. In the extreme version of this, $I_{N-n,n}(\lambda)$ is
zero except in the neighborhood of $\lambda =0$. The integral in
\eq{S_m,n=U_BU_F int I(lambda)} therefore becomes
$(^N_n)\overline{\Pi}(0)\propto \Pi(0)$, where $\overline{\Pi}(0)$ is the
projection operator which projects onto the zero eigenvalue subspace, and so
$S_{N-n,n}=U_{B}[(N-n)\tau ]U_{ F}(n\tau)(^N_n)\overline{\Pi}(0)$. But
according to the nonzero eigenvalue condition in \eq{Pi(0)|psi_0>=0}, the
origin state of the universe has no overlap with any zero-eigenstate state of
the commutator and so $S_{N-n,n}\ket{\psi_{0}}=0$. In other words, the
corresponding paths tend to undergo complete \textit{destructive
interference} in this regime, and make negligible contribution to the sum on
the right side of \eq{N step biev = S_N-n,n}.

The physical implications of these two limiting cases become apparent on considering both
forwards and backwards aggregated intervals $t_{F} =n\tau $ and $t_{B} =(N-n)\tau $ to be
very much larger than the Planck time $\tau $ , i.e. $n,(N-n)\gg 1$. In this regime the
width of the peaks of $I_{N-n,n} (\lambda)$ can be approximated as
\[
    W_{N-n,n} =\frac{\sqrt {24}}{\sqrt {\tau ^{4}n(N-n)N}
   }[1+{\cal O}(\frac{1}{N})]\approx \frac{\sqrt {24}}{\tau ^{1/2}\sqrt {t_{F}
    t_{B} t_{\rm tot}}}
\]
where $t_{\rm tot}=t_{F}+t_{B} =N\tau $ is the total time. Complete destructive
interference occurs for $W_{N-n,n}\ll\lambda_{\rm SD} $ which implies that
$(t_{F}t_{B}t_{\rm tot})^{1/3}\gg\tau^{-1/3}(\lambda_{\rm SD})^{-2/3}$. For $t_{F}\sim
t_{B}$ this implies
\[
    t_{\rm tot} \gg\frac{1}{\tau ^{1/3}(\lambda _{{\rm SD}}
   )^{2/3}}\approx f^{-1/3}10^{-23}{s}\quad .
\]
The destructive interference means that the corresponding set of paths do not contribute to
the sum on the right side of \eq{N step biev = S_N-n,n}; these paths are effectively
``pruned'' from the set possible paths that the universe can take.

It is the condition $t_{F}\sim t_{B}\gg\tau $ that is responsible for this destructive
interference. So to find paths that survive the interference we need to consider times
$t_{F}\ll t_{B}$ or $t_{B}\ll t_{F}$ for total times $t_{\rm tot} =N\tau $ which are at
least measurable. For example, consider total times at the current resolution limit in
time metrology \cite{Rosenband} in which case $t_{{\rm tot}}\approx 10^{-17}{s}$ and so
$N\approx 10^{27}$. Clearly there is no destructive interference for the cases $n=0$ or
$n=N$ (because there is only one path in each case). Also for $n\approx 1$ and $n\approx
N-1$ the estimated width of the peaks of $I_{N-n,n}(\lambda)$ are given by
\begin{equation}
    \label{W_n,m for limit of time metrology}
    W_{N-n,n} =\frac{\sqrt {24}}{\tau ^{2}N}[1+{\cal O}(\frac{1}{N})]\quad .
\end{equation}
In these cases $W_{N-n,n} \approx 10^{61}{s}^{-2}\gg\lambda_{\rm SD} $ which implies
\textit{negligible} destructive interference. This means that the sum in \eq{N step biev
= S_N-n,n} includes a number of paths, however, the paths are characterized by one of
either $t_{F} $ or $t_{B} $ being negligible compared to the total time of $10^{-17}{s}$.

\begin{figure}
\begin{minipage}{45mm}
\vspace{-1cm}\caption{\label{fig: tree with interference}The ``pruning'' of
the binary tree in Fig.~\ref{fig:tree with no interference} by destructive
interference for the case $\Hf\ne\Hb$ and $N\gg 1$.  The relative weighting
of each path segment is depicted by the greyness of the corresponding arrow
with black and white representing maximal and minimal weighting,
respectively: (a) shows an expanded view of the detail near the root node and
(b) shows the whole tree on a much coarser scale.}
\end{minipage}
\hspace{5mm}
\begin{minipage}{100mm}
  \includegraphics[width=100mm]{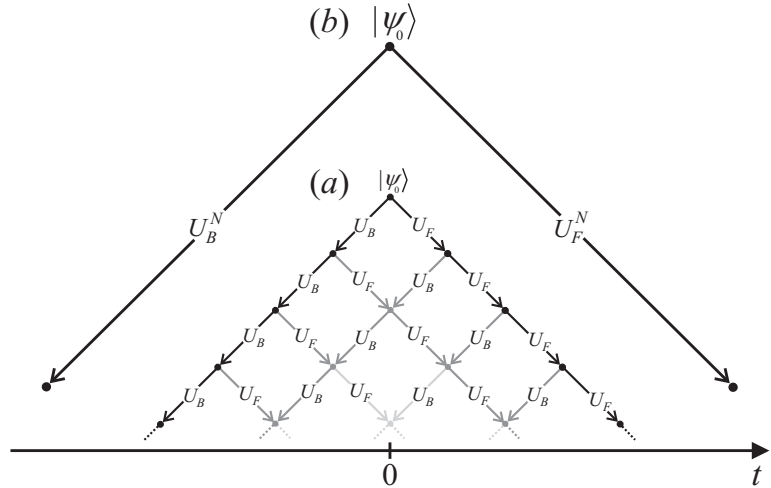}
\end{minipage}
\end{figure}

Finally consider the value of $t_{\rm tot}$ for which there is complete destructive
interference for the sets of paths represented by $\bra{\phi}S_{N-n,n}\ket{\psi_{0}}$ for
all values of $n$ except for $n=0$ and $n=N$. In other words there is complete
destructive interference for $n=1,2,\ldots (N-1)$ and so \eq{N step biev = S_N-n,n} is
simply $\ket{\Psi(N\tau)}=S_{N,0}\ket{\psi_{0}}+S_{0,N} \ket{\psi_{0}}$. We only need to
find the value of $t_{\rm tot}$ for the case $n=1$ as this also ensures the destructive
interference in the remaining cases. Thus we want to know the value of $t_{\rm tot}$ that
satisfies $W_{N-1,1}\ll\lambda_{\rm SD}$. The width $W_{N-1,1}$ is approximated by
\eq{W_n,m for limit of time metrology} and so we have
$\sqrt{24}/(\tau^{2}N)\ll\lambda_{\rm SD} $ which implies $t_{\rm tot} \gg
f^{-1/2}10^{-13}{s}$. In this regime, the two paths that survive the destructive
interference represent exclusively forwards and exclusively backwards evolution.  Figure
\ref{fig: tree with interference} shows the effect of the destructive interference on the
binary tree representation of the evolution.

In summary, for total times $t=N\tau >10^{-17}{s}$ \eq{N step biev = S_N-n,n}
can be replaced with $\ket{\Psi(t)}=[\sum_{n\approx 0}{S_{N-n,n}}
+\sum_{n\approx N}{S_{N-n,n}}]\ket{\psi_{0}}$ and the integral in
\eq{S_m,n=U_BU_F int I(lambda)} is effectively the identity operator. We note
that for $n\approx 0$, \eq{S_m,n=U_BU_F int I(lambda)} then implies that
$S_{N-n,n}=U_{{B}}(N\tau)+{\cal O}(\tau)$. Similarly, it can be shown using
the symmetry property \eq{symmetry I_m,n=I_n,m} that $S_{N-n,n}=U_{
F}(N\tau)+{\cal O}(\tau)$ for $n\approx N$. On ignoring the terms of order
$\tau$ we arrive at a key result of this paper, the \textit{bievolution
equation of motion}
\begin{equation}
    \label{bievolution equation Psi_Bi}
    \ket{\Psi (t)}=\left[ {U_{B} (t)+U_{F} (t)}
    \right]\,\ket{\psi _{0}}\  .
\end{equation}
The term ``bievolution'' here refers to the dual evolution generated by the two different
Hamiltonians. The approximations made in deriving this equation become exact in the limit
$t\gg f^{-1/2}10^{-13}{s}$.

\section{Unidirectionality of time and standard quantum theory \label{sec:unidirectionality}}

\begin{figure}
\begin{minipage}{45mm}
\vspace{-1.5cm}\caption{\label{fig:timeline} The actions of $U_{F}(t)$ and
$U_{B}(t)$ on various states of the universe in relation to clock time $t$.
Diagram (a) represents the actions of $U_{F}(\tau)$ and $U_{B}(\tau)$ on an
arbitrary state $\ket{\psi}$ at clock time $t_0$.  Diagrams (b) and (c)
represent the description of the evolution of the universe by an observer who
finds evidence that the Hamiltonian is $H_F$ or $H_B$, respectively. }
\end{minipage}
\hspace{5mm}
\begin{minipage}{100mm}
\begin{center}
  \includegraphics[width=80mm]{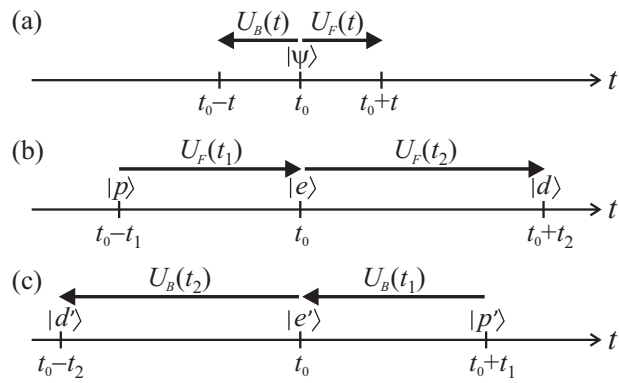}
\end{center}
\end{minipage}
\end{figure}

Consider a situation where the Hamiltonians ${H}_{F}$ and ${H}_{B}$ leave distinguishable
evidence in the state of the universe. For example, imagine that the universe's state at
(total) time $t_{1}$ is
\[
    \ket{\Psi(t_{1})}=\ket{\phi}+\ket{\theta}
\]
where $\ket{\phi}=U_{B}(t_{1})\ket{\psi_{0}}$ and $\ket{\theta}=U_{{
F}}(t_{1})\ket{\psi_{0}}$.  Let $U_{\mu}(t_1+a)=U_{\mu}(a)U_{\mu}(t_1)$ and
$U_{\mu}(a)\ket{x}=\ket{x_{\mu}}$ where $\mu$ is either ${F}$ or ${B}$, $x$ is either
$\phi $ or $\theta $, and $a$ is a suitable fixed time interval. Then, according to
\eq{bievolution equation Psi_Bi}
\beqan
    \ket{\Psi (t_{1}+a)}&=&U_{B}(t_1+a)\ket{\psi_0}+U_{F}(t_1+a)\ket{\psi_0}\\
     &=&U_{B}(a)U_{B}(t_1)\ket{\psi_0}+U_{F}(a)U_{F}(t_1)\ket{\psi_0}\\
    &=&U_{B} (a)\ket{\phi}+U_{F}(a)\ket{\theta}\\
    &=&\ket{\phi _{B}}+\ket{\theta _{F}}.
\eeqan
The Hamiltonians leave {\it distinguishable} evidence in the state of the universe if they
generate orthogonal states in the sense that $\ip{{\theta_{B} }|{\theta_{F}}}=\ip{{\phi
_{B}}|{\phi_{F}}}=0$. We assume this to be the case. Next, let this process occur twice such
that $U_{\nu}(a)U_{\mu}(a)\ket{x}=\ket{x_{\mu,\nu}}$ and
$U_{\mu}(t_1+2a)=U_{\mu}(a)U_{\mu}(t_1+a)$ where $\mu$ and $\nu$ are each either ${F}$ or
${B}$. We find, again by \eq{bievolution equation Psi_Bi}, that
\newpage
\beqa
    \ket{\Psi (t_{1} +2a)}
    &=&U_{B}(t_1+2a)\ket{\psi_0}+U_{F}(t_1+2a)\ket{\psi_0}\non\\
    &=&U_{B}(a)U_{B}(t_1+a)\ket{\psi_0}+U_{F}(a)U_{F}(t_1+a)\ket{\psi_0}\non\\
    &=&U_{B} (a)\ket{\phi _{B}}+U_{F} (a)\ket{\theta _{F}}\non\\
    &=&\ket{\phi _{B,B}}+\ket{\theta _{F,F}}
    \label{corr evidence}
\eeqa
which shows that \textit{corroborating} evidence is left in each term of the
superposition. Recall that the opposite signs in the exponents of the
definitions of $U_{F}$ and $U_{B}$ in Eqs.~(\ref{defn UF}) and (\ref{defn
UB}) imply that they represent time evolution in opposing directions of time,
as illustrated by Fig.~\ref{fig:timeline}(a). We conclude that the
expressions $U_{ F}(t)\ket{\psi}$ and $U_{B}(t)\ket{\psi}$ represent the
universe evolving in opposite directions of time and in each case the
evolution leaves corroborating evidence of the associated Hamiltonian in the
state of the universe.

To interpret what these results mean, consider an observer within the universe who performs
measurements to determine the nature of the Hamiltonian.  According to \eq{corr evidence} the
observer would be in a superposition of two states, one representing the observer consistently
obtaining physical evidence that the Hamiltonian is $H_{F}$, and the other representing the
observer consistently obtaining physical evidence that the Hamiltonian is $H_{{B}}$.  This
means that each term on the right side of the bievolution equation \eq{bievolution equation
Psi_Bi} represents the observer having access to \textit{only one version} of the Hamiltonian.
It follows that the observer would describe the universe as evolving according to either
\begin{equation}
    \ket{\Psi (t)}=U_{F}(t)\,\ket{\psi _{0}}
    \label{schrod F}
\end{equation}
or
\begin{equation}
    \ket{\Psi (t)}=U_{B}(t)\,\ket{\psi _{0}}
    \label{schrod B}
\end{equation}
depending on whether the observer has evidence that the Hamiltonian is $H_F$ or $H_B$,
respectively. Each of \eq{schrod F} and \eq{schrod B} is the solution of the conventional
Schr\"{o}dinger equation for the corresponding version of the Hamiltonian.  The observer's
description would therefore be in accord with our own experience of the dynamics of the
universe.

It is worth remarking that there is no inherent preferred direction of time
in conventional quantum mechanics. Indeed, in conventional quantum theory the
Hamiltonian is the generator of time translations in either direction of
time. Here, however, each of \eq{schrod F} and \eq{schrod B} is associated
with a particular direction of time evolution in the sense that $U_{
F}(\tau)$ evolves the state of the universe along the positive$-t$
(``forward'') direction and $U_{B}(\tau)$ evolves the state of the universe
along the negative$-t$ (``backward'') direction, as illustrated in
Figs.~\ref{fig:timeline}(b) and \ref{fig:timeline}(c).  This association
stems from $H_F$ being the generator of time translations
\textit{exclusively} in the positive$-t$ direction and $H_B$ being the
generator of time translations \textit{exclusively} in the negative$-t$
direction in accord with Wigner's time reversal operation \cite{Wigner} as
discussed in Section 2. The association has important consequences for the
order that events occur in time.  For example, consider the case where the
observer finds evidence that the version of the Hamiltonian is $H_F$ as shown
in Fig.~\ref{fig:timeline}(b). Time evolution in this case is in the
positive$-t$ direction. Imagine that a particular atom is undergoing
spontaneous radiative decay at $t=t_0$ and let $\ket{e}$ represent the
universe with the atom in an excited energy eigenstate. The state
$\ket{d}=U_F(t_2)\ket{e}$ for an appropriate value of $t_2$ would represent
the universe with the atom having decayed to a lower energy eigenstate along
with its emitted radiation, say. The state $\ket{p}$, where
$\ket{e}=U_F(t_1)\ket{p}$, could represent a stage in the preparation of the
excited atom. The order that the preparation stage, excited atom and decay
atom appear in time is dictated by the direction of the time evolution. The
converse case where the observer finds evidence that the Hamiltonian is $H_B$
is illustrated in Fig.~\ref{fig:timeline}(c) for which $\ket{d'}$, $\ket{e'}$
and $\ket{p'}$ represent the decayed atom, the excited atom, and the
preparation stage, respectively. The reversed ordering of the events in this
case is due to the opposite direction of time evolution that is associated
with the $H_B$ version of the Hamiltonian.  In both cases the observer would
find physical evidence of a fixed direction of time evolution and that
direction would correspond to the version of the Hamiltonian observed.

We have allowed the universe to take any path through time according to \eq{N
step biev = (U+U)^N}. But provided there are sufficient T violation
processes, destructive interference between paths will result in the universe
evolving according to the bievolution equation \eq{bievolution equation
Psi_Bi}. We have shown that in this case an observer within the universe
would find physical evidence of only one version of the Hamiltonian. As a
consequence, the observer would describe the universe as evolving according
to the Schr\"{o}dinger equation in the direction of time that corresponds to
the observed version of the Hamiltonian. We have therefore demonstrated the
\textit{consistency between the bievolution equation and conventional quantum
theory for an observer within the universe} and how the
\textit{unidirectional nature of time arises from the perspective of the
observer}.

\section{Eigenstates of $i[{H}_{F}, {H}_{B}]$ with zero eigenvalues} \label{sec: relax
condt nonzero eigenvalues}

We now temporarily relax the nonzero eigenvalue condition in
\eq{Pi(0)|psi_0>=0} and set the origin state to be an eigenstate of the
commutator $i[{H}_{F}, {H}_{B} ]$ with eigenvalue zero, i.e. we set
\[
    \overline{\Pi}(0)\ket{\psi_{0}}=\ket{\psi _{0}}\ .
\]
As previously mentioned, the integral in \eq{S_m,n=U_BU_F int I(lambda)} in this case can
effectively be replaced with $I_{m,n} (0)\rho (0)\Pi (0)=(^{n+m}_{\ \
n})\overline{\Pi}(0)$ and so \eq{N step biev = S_N-n,n} becomes
\[
     \ket{\Psi (N\tau)}=\sum\limits_{n=0}^N U_{B}[(N-n)\tau]U_{F}(n\tau)\left(^N_n\right)\ket{\psi _{0}}
\]
which represents all $2^{N}$ possible paths through time. We now reapply the analysis of the
previous section for the processes $\Uf(a)$ and $\Ub(a)$ with $t_1=N\tau$ and $a=k\tau$. There
is no destructive interference in this case, and so the state of the universe at the total
time $t_1+2a$ is, according to \eq{N step biev = (U+U)^N},
\beqan
     \ket{\Psi (t_1+2a)}=[U_{B} (\tau)+U_{F} (\tau)]^{2k}
     \left[\sum\limits_{n=0}^N U_{B}[(N-n)\tau]U_{F}(n\tau)\left(^N_n\right)\right]\ket{\psi _{0}}
\eeqan
which represents $2^{N+2k}$ paths through time. These paths include the operator combinations
$U_{F}(a)U_{F}(a)$, $U_{F}(a)U_{B}(a)$, $U_{B}(a)U_{F}(a)$, $U_{ B}(a)U_{B}(a)$ (among many
others) and so any distinguishable evidence of the Hamiltonians left in the state of the
universe is not necessarily corroborated over repetitions of the process. This is not in
accord with present experimental evidence. Moreover, consider the origin state of the most
general form
\[
    \ket{\psi_0}=\ket{\psi_0^\perp}+\ket{\psi_0^\|}
\]
where $\overline{\Pi}(0)\ket{\psi_0^\perp}=0$ and
$\overline{\Pi}(0)\ket{\psi_0^\|}=\ket{\psi_0^\|}$. For this case the bievolution
equation (\ref{bievolution equation Psi_Bi}) is replaced with
\beq
    \label{eq:bievolution with [H,H]=0 states}
    \ket{\Psi (t)}=\left[ {U_{B} (t)+U_{F} (t)}
    \right]\,\ket{\psi_{0}^\perp}
    +\sum\limits_{n=0}^N U_{B}[(N-n)\tau]U_{F}(n\tau)\left(^N_n\right)\ket{\psi_{0}^\|}
\eeq
where $t=N\tau$.  However, as we have already seen, experimental evidence
suggests that the universe is not following the corresponding set of paths
represented by the last term on the right side. In order for our system to
provide a physical model of the {\it observable} universe, it is therefore
sufficient to only consider origin states that satisfy the nonzero eigenvalue
condition in \eq{Pi(0)|psi_0>=0}. We assume this to be the case in the
remainder of this work.

\section{Schr\"{o}dinger's equation for bievolution}

We can construct the differential form of the bievolution equation of motion as follows.
Increasing $t$ in \eq{bievolution equation Psi_Bi} by a relatively small time interval
$\delta t$ yields $\ket{\Psi(t+\delta t)}=\left[ {U_{B}(\delta t)U_{B}(t)+U_{F}(\delta
t)U_{F}(t)} \right]\ket{\psi_{0}}$ and so the rate of change of the state is given by
\[
     \frac{\delta \ket{\Psi (t)}}{\delta t}=\left[
     {i{H}_{B} U_{B} (t)-i{H}_{F} U_{F} (t)}
     \right]\,\ket{\psi _{0}}+{\cal O}(\delta t)
\]
where $\delta\ket{\Psi(t)}=\ket{\Psi(t+\delta t)}-\ket{\Psi(t)}$. Taking the
limit $\delta t\to \tau $ and ignoring a term of order $\tau $ gives another
key result, the \textit{Schr\"{o}dinger equation for bievolution}
\[
     \frac{d\ket{\Psi (t)}}{dt}=\frac{d\ket{\psi _{{
     F}} (t)}}{dt}-\frac{d\ket{\psi _{B} (t)}}{dt}
\]
where $(d/dt)\ket{\psi _{\mu}(t)}=-i{H}_{\mu}\ket{\psi_{\mu}(t)}$ and $\ket{\psi _{\mu
}(t)}=U_{\mu}(t)\ket{\psi_{0}}$ for $\mu={F}$ or ${
B}$.

\section{Discussion}

To gauge the full impact of T violation processes, it is instructive to compare the above
analysis with a universe which obeys T invariance. In this case ${H}_{F}={H}_{{B}}={H}$ and it
can be shown that
\beq
    S_{N-n,n} =\exp [i(N-2n)\tau{H}]\left(^N_n\right)
    \label{eq: S_N,n for H_F=H_B}
\eeq
which is never zero and so all possible paths are included in \eq{N step biev = S_N-n,n}. The
direction of time is ambiguous as there is no physical evidence of any kind to single out one
direction over the other. The fact that this ambiguity is removed for a universe with T
violation processes leads directly to the proposition that \textit{T violation processes are
responsible for the phenomenological unidirectionality of time that we observe in the
universe.}

Just as important is the implicit assertion here that {\it in the absence of
a unidirectional nature, time is randomly directed at any given instant.} We
allowed for the possibility of a random direction when we added amplitudes in
\eq{add amplitudes}. In hindsight, the analysis of the preceding sections can
be seen to be a study of the dichotomy between randomly directed and
unidirectional time evolution. In this context the important question about
the unidirectional nature of time is not how one direction of time is chosen
over the other, but rather how each direction is maintained consistently from
one instant to the next. This is one of the underlying questions that is
being addressed in this article, with T violation providing a possible
mechanism as the answer.

We can also elaborate on what the time reversal operation means in the
presence of T violating processes. The bievolution equation \eq{bievolution
equation Psi_Bi} shows that as the value of $t$ increases, the states $U_{F}
(t)\ket{\psi _{0}}$ and $U_{B} (t)\ket{\psi _{0}}$ in each branch of the
superposition trace out respective trajectories in the state space.
Decreasing values of $t$ then simply ``back tracks'' the states along their
respective trajectories in a consistent manner, essentially tracking back
over what might be called their ``histories''. In contrast, the time reversal
operation essentially interchanges $\Uf(t)$ and $\Ub(t)$ in \eq{bievolution
equation Psi_Bi}.  This implies that the time reversal operation refers to
\textit{switching between the branches of the bievolution equation} as
opposed to tracking back over histories.

These results significantly elevate the importance of the matter-antimatter
arrow in the study of the nature of time.  The underlying T violation is
based on the empirical evidence that different versions of the Hamiltonian
operate for different directions of time evolution. Moreover, as discussed in
Section \ref{sec:unidirectionality}, each direction of time can be uniquely
identified by physical evidence left by the Hamiltonian. This is quite
different to other arrows of time. In particular, the thermodynamic arrow
does not distinguish between the different directions of time evolution in
this way because entropy increases in the direction of time evolution {\it
regardless of the direction}. This suggests that T violation plays a far more
significant role in determining the nature of time than previously imagined.

We should mention that the physical implications of T violation described
here depend the magnitudes of the eigenvalues $\lambda $. Our estimate of
these eigenvalues depends on the number of particles involved in T violations
processes and this has been fixed only at $f 10^{80}$ where $f<1$ is a
parameter of undetermined value. However, our analysis does not depend
critically on the value of $f$ and, indeed, a large range of its values
support the key results. Also, our analysis has focused on relatively large
time scales in order to draw out the main consequences of T violation for the
unidirectionality of time. In doing so we have ignored terms of order $\tau$,
the Planck time. These terms imply that the directionality of time has finer
details for relatively small time intervals. The elaboration of these details
is, however, beyond the scope of this article.

There are potentially important experimental implications of the analysis
presented here. A sufficiently small value of $f$ would produce incomplete
destructive interference and a deviation from the dynamics described by the
bievolution equation \eq{bievolution equation Psi_Bi}.  This would be
manifest in a lower bound to the accuracy of time metrology. A more extreme
case occurs in the very early universe during the inflation and
radiation-dominated periods \cite{bennett,reheating}.  Baryogenesis would not
have yet occurred and CP and T violation would have been relatively rare
events. This implies that the value of $f$ would have been negligible during
these periods and so the direction of time evolution would have been
uncertain. This has potentially important ramifications for cosmological
models and their testing against observational data of the cosmic microwave
background radiation \cite{bennett}.

The analysis presented here may also have important consequences for quantum
gravity. Consider again the situation where $H_F=H_B=H$.  Instead of
expanding \eq{N step biev = (U+U)^N} into terms of the kind in \eq{eq: S_N,n
for H_F=H_B} we write $U_F+U_B=2\cos(H\tau)$ and so
\beq
   \ket{\Psi(N\tau)}=2^N\cos^N(H\tau)\ket{\psi_0}\ .
\eeq
Let the eigenbasis of $H$ be given by
\beq
  H\ket{E_n,\lambda}=E_n\ket{E_n,\lambda}\ ,
\eeq
where $E_n$ are the energy eigenvalues with $E_0=0$ and $\lambda$ indexes
different eigenstates in degenerate manifolds. Expanding $\ket{\psi_0}$ in
this basis gives
\beq
   \ket{\Psi(N\tau)}=2^N\sum_{n,\lambda} c_{n,\lambda}\cos^N(E_n\tau)\ket{E_n,\lambda}
\eeq
where $c_{n,\lambda}=\ip{E_n,\lambda|\psi_0}$. Let $\ket{\psi_0}$ be a state
of bounded energy to the extent that $c_{n,\lambda}=0$ for $|E_n| \ge
1/\tau$. Here $1/\tau$ is the Planck energy in units where $\hbar=1$. This
bound would be reasonable for a very young and very small universe.  In this
case, provided $c_{0,\lambda}\ne 0$ for some values of $\lambda$, the state
$\ket{\Psi(N\tau)}$ becomes proportional to $\sum_\lambda
c_{0,\lambda}\ket{E_0,\lambda}$ as $N\to\infty$. Hence in the limit as
$N\to\infty$
\beq
   H\ket{\Psi(N\tau)}=0\ .
   \label{eq: H|Psi>=0}
\eeq
According to \eq{N step biev = (U+U)^N}, $\ket{\Psi(N\tau)}$ is a superposition of states each
of which represents a net time ranging from $-N\tau$ to $+N\tau$.  It therefore represents the
whole {\it history} of the universe. \eq{eq: H|Psi>=0} is analogous to the topological
invariance condition discussed by Misner \cite{Misner} in relation to the quantization of
gravity. Topological invariance is needed in quantum gravity to maintain the invariance of
physical quantities to transformations of spacetime manifolds. To make a direct connection
with Misner's work, we let $H$ represent the Hamiltonian $H(x)$ for gravitational and matter
fields defined on a four dimensional spacetime manifold $x$, $\{\sigma_k\}$ represent a set of
spacelike hypersurfaces indexed by integer $k$ and separated by timelike intervals of $\tau$,
$U_F(n\tau)\ket{\psi_0}$ represent the state functional $\psi_{\sigma_n}$ and
$U_B(n\tau)\ket{\psi_0}$ represent the state functional $\psi_{\sigma_{-n}}$ for positive
integer $n$, where $\psi_\sigma$ is a functional of the gravitational and matter fields at the
hypersurface $\sigma$. In this representation $\ket{\Psi(N\tau)}$ represents a state
functional which is a superposition of $\psi_{\sigma_{n}}$ for $n=-N$ to $N$. \eq{eq:
H|Psi>=0} then represents the result that {\it topological invariant states are attractors of
time symmetric evolution}. The zero energy eigenstate implies that the positive energy of
matter fields is balanced by the negative energy of the gravitational potential energy
\cite{Tolman,Tryon,Hartle}.

This means that \eq{eq: H|Psi>=0} corresponds to the Hamiltonian constraint
of the Wheeler-DeWitt equation \cite{DeWitt,Alvarez}. The Wheeler-DeWitt
equation represents the entire history of the universe with no bias towards
either direction of time. Here time symmetric evolution is unbiased in the
same way in the sense that it represents evolution without a fixed direction
of time and the state in \eq{eq: H|Psi>=0} represent the entire history of
the universe. These common features suggest that there is an elemental
relationship between the Wheeler-DeWitt description and the present analysis.
The present analysis may therefore offer a means of incorporating the two
versions of the matter Hamiltonian in a Wheeler-DeWitt description of the
universe for situations where $H_F\ne H_B$. Moreover, as the present analysis
shows how the phenomenological unidirectionality of time arises when $H_F\ne
H_B$, it may also shed fresh light on the problem of how time is established
in the timeless Wheeler-DeWitt description.

In conclusion, the physical significance of the relatively weak and rare T violation processes
exhibited by mesons has been a puzzle for more than four decades. Despite representing a
fundamental time asymmetry, these processes have been regarded as having an insignificant
affect on the physical nature of time. This has been due in part to the lack of a formalism
for accommodating their time asymmetry in a single dynamical law. We have removed this
obstacle by deriving, from a first principles analysis based on Feynman's sum over histories,
a general method for incorporating both versions of the Hamiltonian for a T violating process,
one for forwards and the other for backwards evolution, in a single dynamical equation of
motion. Moreover we have shown that these processes can affect the time evolution of the
universe on a grand scale. Indeed, we have shown that T violation provides, at least in
principle, a physical mechanism for the phenomenological unidirectionality of time. Finally,
the analysis presented here may also have important ramifications for time metrology, quantum
gravity and the early development of the universe. At the very least it gives a clue to the
quantum nature of time itself.

\begin{acknowledgements}
The author thanks Prof. D.\ T.\ Pegg for sharing his wealth of knowledge on
the nature of time and for the generous advice he offered in the development
of this article, Prof. S.\ M.\ Barnett for his encouragement and help with
kaon dynamics and Prof. H.\ M.\ Wiseman for helpful discussions.
\end{acknowledgements}

\appendix

\renewcommand{\thesection}{\large Appendix \Alph{section}}

\section{\large\kern-1mm: Reordering factors of $S_{m,n}$}\label{app-reordering}
\renewcommand{\theequation}{\Alph{section}.\arabic{equation}}\setcounter{equation}{0}

The operator $S_{m,n}$ is defined recursively in \eq{defn S_m,n} as
\beq
     S_{m,n}=\sum_{k=0}^{m} S_{m-k,n-1} U_{F}(\tau)U_{B}(k\tau)
     \label{recursive}
\eeq
with $S_{m,0}=U_{B}(m\tau)$ and so
\beqa
   S_{m,1}&=&\sum_{k=0}^{m} S_{m-k,0}U_{F}(\tau)U_{B}(k\tau)
   =\sum_{k=0}^{m}U_{B}[(m-k)\tau] U_{F}(\tau)U_{B}(k\tau)\ .
\eeqa
The Zassenhaus formula \cite{Suzuki}
\[
    e^{A\epsilon}e^{B\epsilon}e^{-\frac{1}{2}\epsilon^2[A,B]}e^{{\cal O}(\epsilon^3)}
    =e^{B\epsilon}e^{A\epsilon}e^{-\frac{1}{2}\epsilon^2[B,A]}e^{{\cal O}(\epsilon^3)}
\]
(which is related to the Baker-Campbell-Hausdorff formula) where $A$ and $B$ are
arbitrary operators and $\epsilon$ is a small parameter, can be written as
\[
    e^{A\epsilon}e^{B\epsilon}
    =e^{B\epsilon}e^{A\epsilon}e^{\epsilon^2[A,B]}e^{{\cal O}(\epsilon^3)}
\]
which shows that
\beq
    e^{-ij\tau\Hf}e^{ik\tau\Hb }
    =e^{ik\tau\Hb }e^{-ij\tau\Hf}e^{jk\tau^2[\Hf,\Hb]}e^{{\cal O}(\tau^3)}\ .
\eeq
In terms of the operators $\Uf(\tau)=e^{-i\tau\Hf}$ and $\Ub(\tau)=e^{i\tau\Hb}$ this
result reads
\beq
     \Uf(j\tau)\Ub(k\tau)
    =\Ub(k\tau)\Uf(j\tau)e^{jk\tau^2[\Hf,\Hb]}e^{{\cal O}(\tau^3)}
\eeq
or equivalently
\beq
         \Ub(k\tau)\Uf(j\tau)
    =\Uf(j\tau)\Ub(k\tau)e^{-jk\tau^2[\Hf,\Hb]}e^{{\cal O}(\tau^3)}\ .
\eeq
Hence
\beqa
     S_{m,n}
     &=&\sum_{k=0}^{m} S_{m-k,n-1}\Uf(\tau)\Ub(k\tau)\non\\
     &=&\sum_{k=0}^{m} S_{m-k,n-1}\Ub(k\tau)\Uf(\tau)e^{k\tau^2[\Hf,\Hb]}e^{{\cal O}(\tau^3)}\non\\
     &=&\sum_{k=0}^{m} \sum_{j=0}^{m-k} S_{m-k-j,n-2}\Uf(\tau)\Ub(j\tau)\Ub(k\tau)\Uf(\tau)e^{k\tau^2[\Hf,\Hb]}e^{{\cal O}(\tau^3)}\non\\
     &=&\sum_{k=0}^{m} \sum_{j=0}^{m-k} S_{m-k-j,n-2}\Uf(\tau)\Ub[(j+k)\tau]\Uf(\tau)e^{k\tau^2[\Hf,\Hb]}e^{{\cal O}(\tau^3)}\non\\
     &=&\sum_{k=0}^{m} \sum_{j=0}^{m-k} S_{m-k-j,n-2}\Uf(2\tau)\Ub[(j+k)\tau]e^{-j\tau^2[\Hf,\Hb]}e^{{\cal O}(\tau^3)}\non\\
     &=&\sum_{k=0}^{m} \sum_{j=0}^{m-k} S_{m-k-j,n-2}\Ub[(j+k)\tau]\Uf(2\tau)e^{(j+2k)\tau^2[\Hf,\Hb]}e^{{\cal
     O}(\tau^3)}\ .
\eeqa
Setting $\ell=j+k$ and rearranging the order the terms are summed to eliminate $j$ then
yields
\beqa
     S_{m,n}
     &=&\sum_{\ell=0}^{m} \sum_{k=0}^{\ell} S_{m-\ell,n-2}\Ub(\ell\tau)\Uf(2\tau)e^{(\ell+k)\tau^2[\Hf,\Hb]}e^{{\cal
     O}(\tau^3)}\non\\
     &=&\sum_{\ell=0}^{m} S_{m-\ell,n-2}\Ub(\ell\tau)\Uf(2\tau)e^{\ell\tau^2[\Hf,\Hb]}\sum_{k=0}^{\ell}e^{k\tau^2[\Hf,\Hb]} e^{{\cal
     O}(\tau^3)}\non\\
     &=&\sum_{\ell=0}^{m} \sum_{r=0}^{m-\ell} S_{m-\ell-r,n-3}\Uf(\tau)\Ub[(\ell+r)\tau]\Uf(2\tau)e^{\ell\tau^2[\Hf,\Hb]}\sum_{k=0}^{\ell}e^{k\tau^2[\Hf,\Hb]} e^{{\cal
     O}(\tau^3)}\non\\
     &=&\sum_{\ell=0}^{m} \sum_{r=0}^{m-\ell} S_{m-\ell-r,n-3}\Ub[(\ell+r)\tau]\Uf(3\tau)e^{(r+2\ell)\tau^2[\Hf,\Hb]}\sum_{k=0}^{\ell}e^{k\tau^2[\Hf,\Hb]} e^{{\cal
     O}(\tau^3)}\ .
\eeqa
Rearranging as before but with $j=\ell+r$ and eliminating $r$ yields
\beqa
     S_{m,n}
     &=&\sum_{j=0}^{m} \sum_{\ell=0}^{j} S_{m-j,n-3}\Ub(j\tau)\Uf(3\tau)e^{(j+\ell)\tau^2[\Hf,\Hb]}\sum_{k=0}^{\ell}e^{k\tau^2[\Hf,\Hb]} e^{{\cal
     O}(\tau^3)}\non\\
     &=&\sum_{j=0}^{m} S_{m-j,n-3}\Ub(j\tau)\Uf(3\tau)e^{j\tau^2[\Hf,\Hb]} \sum_{\ell=0}^{j}\sum_{k=0}^{\ell}e^{(\ell+k)\tau^2[\Hf,\Hb]} e^{{\cal
     O}(\tau^3)}\ .
\eeqa
Continuing in this way gives, after $w$ uses of the recursive relation \eq{recursive},
\beqa
     S_{m,n}
     &=&\sum_{v=0}^{m} S_{m-v,n-w}\Ub(v\tau)\Uf(w\tau)e^{v\tau^2[\Hf,\Hb]}\sum_{r=0}^{v}\cdots \sum_{\ell=0}^{s}\sum_{k=0}^{\ell}e^{(r+\cdots+\ell+k)\tau^2[\Hf,\Hb]} e^{{\cal
     O}(\tau^3)}
\eeqa
in which there are $w$ summations.  Letting $w=n$ then yields
\beqa
     S_{m,n}
     &=&\sum_{v=0}^{m} S_{m-v,0}\Ub(v\tau)\Uf(n\tau)e^{v\tau^2[\Hf,\Hb]}\sum_{r=0}^{v}\cdots \sum_{\ell=0}^{s}\sum_{k=0}^{\ell}e^{(r+\cdots+\ell+k)\tau^2[\Hf,\Hb]} e^{{\cal
     O}(\tau^3)}\non\\
     &=&\sum_{v=0}^{m} \Ub(m\tau)\Uf(n\tau)e^{v\tau^2[\Hf,\Hb]}\sum_{r=0}^{v}\cdots \sum_{\ell=0}^{s}\sum_{k=0}^{\ell}e^{(r+\cdots+\ell+k)\tau^2[\Hf,\Hb]} e^{{\cal
     O}(\tau^3)}\non\\
     &=&\Ub(m\tau)\Uf(n\tau)\sum_{v=0}^{m} \sum_{r=0}^{v}\cdots \sum_{\ell=0}^{s}\sum_{k=0}^{\ell}e^{(v+r+\cdots+\ell+k)\tau^2[\Hf,\Hb]} e^{{\cal
     O}(\tau^3)}\ .
\eeqa
Hence we arrive at \eq{S_m,n=U_BU_F sum commutator}
\beq
     S_{m,n}
     =\Ub(m\tau)\Uf(n\tau)\sum_{v=0}^{m}\cdots \sum_{\ell=0}^{s}\sum_{k=0}^{\ell}e^{(v+\cdots+\ell+k)\tau^2[\Hf,\Hb]} e^{{\cal
     O}(\tau^3)}
     \label{S_Nn_reordered}
\eeq
in which there are $n$ summations.

\section{\large\kern-1mm: Simplifying $I_{m,n}(\lambda)$}\label{app-simplifying}\setcounter{equation}{0}

Consider the following manipulations of {\it nested summations} of the kind
\beqa
      \sum_{j=0}^{w}\sum_{\ell=0}^{j}\sum_{m=0}^{\ell}\cdots\sum_{n=0}^s
      \sum_{k=0}^n e^{i(j+\ell+m+\cdots+n+k)\theta}\ .
\eeqa
We call the summations ``nested'' because the upper limit $n$ of the last summation over
$k$ is the index of the second-last summation and so on.  The nested property means that
summations cannot be evaluated independently of each other.  However it is possible to
reduce the number of nested summations by performing certain operations which we now
describe.

First, by changing the order in which the indices $j$ and $\ell$ are summed, we find
\beqa
      \sum_{j=0}^{w}\sum_{\ell=0}^{j}\sum_{m=0}^{\ell}\cdots\sum_{n=0}^s\sum_{k=0}^n e^{i(j+\ell+m+\cdots+n+k)\theta}
      = \sum_{\ell=0}^{w}\sum_{j=\ell}^{w}\sum_{m=0}^{\ell}\cdots\sum_{n=0}^s\sum_{k=0}^n
      e^{i(j+\ell+m+\cdots+n+k)\theta}\ ,
\eeqa
next, by cyclically interchanging the indices in the order $j\to k\to n\to s\to\cdots\to
m\to\ell\to j$ on the right-hand side, we get
\beqa
     \sum_{j=0}^{w}\sum_{\ell=0}^{j}\sum_{m=0}^{\ell}\cdots\sum_{n=0}^s\sum_{k=0}^n e^{i(j+\ell+m+\cdots+n+k)\theta}
      = \sum_{j=0}^{w}\sum_{k=j}^{w}\sum_{\ell=0}^{j}\sum_{m=0}^{\ell}\cdots\sum_{n=0}^s
      e^{i(k+j+\ell+m+\cdots+n)\theta}\ ,
\eeqa
and finally, bringing the sum over $k$ to the extreme right on the right-hand side gives
\beqa
     \sum_{j=0}^{w}\sum_{\ell=0}^{j}\sum_{m=0}^{\ell}\cdots\sum_{n=0}^s\sum_{k=0}^n e^{i(j+\ell+m+\cdots+n+k)\theta}
      = \sum_{j=0}^{w}\sum_{\ell=0}^{j}\sum_{m=0}^{\ell}\cdots\sum_{n=0}^s\sum_{k=j}^{w}
      e^{i(j+\ell+m+\cdots+n+k)\theta}\ .
\eeqa
We can abbreviate this general summation property as
\beq
     \sum_{j=0}^{w}\cdots\sum_{s=0}^t\sum_{n=0}^s\sum_{k=0}^n e^{i(j+\cdots+s+n+k)\theta}
      = \sum_{j=0}^{w}\cdots\sum_{s=0}^t\sum_{n=0}^s\sum_{k=j}^{w}
      e^{i(j+\cdots+s+n+k)\theta}\ .
      \label{general sum prop}
\eeq
Consider the product
\beqa
     \Big( e^{i\theta}+1\Big)\sum_{n=0}^s\sum_{k=0}^n e^{i(n+k)\theta}
      =e^{i\theta}\sum_{n=0}^s\sum_{k=0}^n e^{i(n+k)\theta}+\sum_{n=0}^s\sum_{k=0}^n
      e^{i(n+k)\theta}\ .
\eeqa
Using \eq{general sum prop} for the two summations in the first term on the right-hand
side gives
\beqa
     \Big( e^{i\theta}+1\Big)\sum_{n=0}^s\sum_{k=0}^n e^{i(n+k)\theta}
      &=&e^{i\theta}\sum_{n=0}^s\sum_{k=n}^s e^{i(n+k)\theta}+\sum_{n=0}^s\sum_{k=0}^n
      e^{i(n+k)\theta}\non\\
      &=&\sum_{n=0}^s\sum_{k=n}^s e^{i(n+k+1)\theta}+\sum_{n=0}^s\sum_{k=0}^n
      e^{i(n+k)\theta}\non\\
      &=&\sum_{n=0}^s\sum_{k=n+1}^{s+1} e^{i(n+k)\theta}+\sum_{n=0}^s\sum_{k=0}^n
      e^{i(n+k)\theta}
      =\sum_{n=0}^s\sum_{k=0}^{s+1} e^{i(n+k)\theta}\ .
      \label{2 sums}
\eeqa
The two {\it nested} summations on the left-hand side have been reduced to two {\it
un-nested} summations on the right-hand side. Similarly, using \eq{general sum prop}
again as well as \eq{2 sums} we find
\beqa
     \hs{-0.7}\Big( e^{i2\theta}\!\!+\!e^{i\theta}\!\!+\!1\Big)\sum_{s=0}^t\sum_{n=0}^s\sum_{k=0}^n e^{i(s+n+k)\theta}
     &=& e^{i2\theta}\sum_{s=0}^t\sum_{n=0}^s\sum_{k=0}^n e^{i(s+n+k)\theta}\non\\
     &&\hspace{2cm}+\Big(e^{i\theta}+1\Big)\sum_{s=0}^t\sum_{n=0}^s\sum_{k=0}^n
     e^{i(s+n+k)\theta}\non\\
     &=& \sum_{s=0}^t\sum_{n=0}^s\sum_{k=s}^t e^{i(s+n+k+2)\theta}
                  +\sum_{s=0}^t\sum_{n=0}^s\sum_{k=0}^{s+1} e^{i(s+n+k)\theta}\non\\
     &=& \sum_{s=0}^t\sum_{n=0}^s\sum_{k=s+2}^{t+2} e^{i(s+n+k)\theta}
                  +\sum_{s=0}^t\sum_{n=0}^s\sum_{k=0}^{s+1} e^{i(s+n+k)\theta}\non\\
     &=& \sum_{s=0}^t\sum_{n=0}^s\sum_{k=0}^{t+2} e^{i(s+n+k)\theta}\ .
\eeqa
Here the {\it three} nested summations on the left-hand side have been reduced to {\it
two} nested summations and one un-nested summation on the right-hand side. This implies
that, for example, $p+1$ nested summations can be reduced to $p$ nested and one un-nested
summations as follows:
\beqa
     \Big(\sum_{r=0}^{p}e^{ir\theta}\Big)\!\!\underbrace{\ \sum_{j=0}^{w}\cdots\sum_{s=0}^t\sum_{n=0}^s\sum_{k=0}^n\
     }\!\!
                        e^{i(j+\cdots+s+n+k)\theta}
      &=& \Big(\!\!\underbrace{\ \sum_{j=0}^{w}\cdots\sum_{s=0}^t\sum_{n=0}^s\ }\!\!
      e^{i(j+\cdots+s+n)\theta}\Big)\sum_{k=0}^{w+p} e^{ik\theta}\ .\non\\
      \mbox{\small $p\!+\!1$ nested sums}\hs{2.4}&&\hs{.5}\mbox{\small $p$ nested sums}
\eeqa
Moreover, by repeating this process once more, the $p$ nested summations of the
right-hand side can be further reduced to $p-1$ nested and one un-nested summations.
Continuing in this way eventually leads to the following result
\beqa
     \prod_{q=0}^{p}\!\Big(\sum_{r=0}^{q}e^{ir\theta}\Big)\!\!\underbrace{\ \sum_{j=0}^{w}\!\cdots\!\sum_{s=0}^t\sum_{n=0}^s\sum_{k=0}^n\ }
                       \!\! e^{i(j+\cdots+s+n+k)\theta}
      &=& \prod_{q=0}^{p-1}\!\Big(\sum_{r=0}^{q}e^{ir\theta}\Big)\Big(\!\underbrace{\ \sum_{j=0}^{w}\!\cdots\!\sum_{s=0}^t\sum_{n=0}^s\
      }\!\! e^{i(j+\cdots+s+n)\theta}\Big)\!\sum_{k=0}^{w+p} e^{ik\theta}\non\\
      \mbox{\small $p\!+\!1$ nested sums}\hs{2.5}
      &&\hs{2.5}\mbox{\small $p$ nested sums} \non\\
      &=& \underbrace{\ \Big(\sum_{j=0}^{w} e^{ij\theta}\Big)\cdots\Big(\sum_{n=0}^{w+p-1} e^{in\theta}\Big)\Big(\sum_{k=0}^{w+p} e^{ik\theta}\Big)\ }\non\\
      &&\hs{1.8}\mbox{\small $p\!+\!1$ un-nested sums} \non\\
      &=&\prod_{q=0}^{p}\Big(\sum_{j=0}^{w+q} e^{ij\theta}\Big)\ .
\eeqa
Thus
\beqa
     \underbrace{\ \sum_{j=0}^{w}\cdots\sum_{s=0}^t\sum_{n=0}^s\sum_{k=0}^n\ }
                        e^{i(j+\cdots+s+n+k)\theta}
      &=&\frac{\prod_{q=0}^{p}\Big(\sum_{j=0}^{w+q}
      e^{ij\theta}\Big)}{\prod_{q=0}^{p}\Big(\sum_{r=0}^{q}e^{ir\theta}\Big)}\ .\\
      \mbox{\small $p\!+\!1$ nested sums}\hs{2.6}&\non
\eeqa
Evaluating the two geometric series on the right-hand side then gives
\beqa
     \underbrace{\ \sum_{j=0}^{w}\cdots\sum_{s=0}^t\sum_{n=0}^s\sum_{k=0}^n\ }
                        e^{i(j+\cdots+s+n+k)\theta}
      &=&\frac{\prod_{q=0}^{p}\Big(1-e^{i(w+q+1)\theta}\Big)}{\prod_{q=0}^{p}\Big(1-e^{i(q+1)\theta}\Big)}
      =\frac{\prod_{q=1}^{p+1}\Big(1-e^{i(w+q)\theta}\Big)}{\prod_{q=1}^{p+1}\Big(1-e^{iq\theta}\Big)}\ .\non\\
      \mbox{\small $p\!+\!1$ nested sums}\hs{2.6}&
\eeqa
Adjusting the number of nested summations of the left-hand side then gives the useful
result that
\beqa
     \underbrace{\ \sum_{j=0}^{w}\cdots\sum_{s=0}^t\sum_{n=0}^s\sum_{k=0}^n\ }
                        e^{i(j+\cdots+s+n+k)\theta}
     &=&\frac{\prod_{q=1}^{p}\Big(1-e^{i(w+q)\theta}\Big)}{\prod_{q=1}^{p}\Big(1-e^{iq\theta}\Big)}\ .\\
      \mbox{\small $p$ nested sums}\hs{2.8}&\non
\eeqa
We can now use this result to simplify the $n$ nested summations in \eq{I(lambda)=big
sum} as follows.  First we note that
\beqa
     I_{m,n}(\lambda)
     =\sum_{v=0}^{m}\cdots
     \sum_{\ell=0}^{s}\sum_{k=0}^{\ell}e^{-i(v+\cdots+\ell+k)\tau^2\lambda}
     &=&\frac{\prod_{q=1}^{n}\Big(1-e^{-i(m+q)\tau^2\lambda}\Big)}{\prod_{q=1}^{n}\Big(1-e^{-iq\tau^2\lambda}\Big)}
\eeqa
and then by redefining the index $q$ in the numerator we arrive at
\beqa
     I_{m,n}(\lambda)
     &=&\frac{\prod_{q=0}^{n-1}\Big(1-e^{-i(n+m-q)\tau^2\lambda}\Big)}{\prod_{q=1}^{n}\Big(1-e^{-iq\tau^2\lambda}\Big)}
\eeqa
which is \eq{I(lambda)=simple} of the paper.

\section{\large\kern-1mm: Estimating eigenvalues}\label{app-eigenvalues}\setcounter{equation}{0}

We estimate the eigenvalues of the commutator $[\Hf^{(1)},\Hb^{(1)}]$ on the state space
of a kaon field mode that represents a single particle. However, to avoid unnecessary
clutter, we will omit the superscript ``$(1)$'' on the single particle Hamiltonians in
the remainder of this section.

It is convenient to invoke the CPT theorem and use the $CP$ operator instead of the $T$
operator as follows
\beqa
   \Hb &=& T\,\Hf  T^{-1}=(CP)^{-1}\,\Hf \, (CP)\ .
   \label{Hb = CP Hf CP}
\eeqa
Consider the matrix representation in the kaon state space where the vectors $(1,0)$ and
$(0,1)$ represent the kaon and anti-kaon states $\ket{K^0}$ and $\ket{\bK{}^0}$,
respectively. In this basis the general form of the $CP$ operator is
\beq
   CP=\mat{cc}{0 & e^{i\eta}\\ e^{i\xi}&0}
\eeq
where $\eta$ and $\xi$ are possible phase angles.  By redefining the complex phase of the
kaon state where $\ket{K^0}$ is multiplied by $e^{i(\eta-\xi)/2}$, the $CP$ operator
becomes
\beq
   CP=\mat{cc}{0 & e^{i\theta}\\ e^{i\theta}&0}=e^{i\theta}\mat{cc}{0 & 1\\ 1&0}
\eeq
where $\theta=\sfc{1}{2}(\eta+\xi)$. (This is a more useful transformation for us here
compared to the usual one, which multiplies $\ket{K^0}$ by $e^{-i\xi}$.) Let $M_{ij}$ be
the matrix elements of $\Hf$ in the kaon state basis where $M_{11}=\ip{K^0|\Hf|K^0}$ etc.
To find the corresponding matrix elements of $\Hf$ we use \eq{Hb = CP Hf CP}, that is
\beqan
   \Hb&=&(CP)^{-1}\,\Hf \, (CP)\\
   &=&e^{-i\theta}\mat{cc}{0 & 1\\1&0}\mat{cc}{M_{11} & M_{12} \\ M_{21} & M_{22}}e^{i\theta}\mat{cc}{0 & 1\\
   1&0}=
   \mat{cc}{M_{22} & M_{21} \\ M_{12} & M_{11}}\ .
\eeqan
Hence we immediately find
\beqan
   \Hf\Hb&=&\mat{cc}{M_{11}M_{22}+(M_{12})^2 & M_{11}(M_{21}+M_{12}) \\ M_{22}(M_{21}+M_{12}) & M_{11}M_{22}+(M_{21})^2}\\
   \Hb\Hf&=&{\mat{cc}{M_{11}M_{22}+(M_{21})^2 & M_{22}(M_{21}+M_{12}) \\ M_{11}(M_{21}+M_{12}) & M_{11}M_{22}+(M_{12})^2}}^{\phantom{M}}\ ,
\eeqan
and so the commutator is
\beqa
   [\Hf,\Hb]=\Hf\Hb-\Hb\Hf&=&\mat{cc}{(M_{12})^2-(M_{21})^2 & (M_{11}-M_{22})(M_{21}+M_{12}) \\
   -(M_{11}-M_{22})(M_{21}+M_{12}) & -[(M_{12})^2-(M_{21})^2]}\non\\
             \label{commutator matrix}
\eeqa
which is in the form
\beqa
   [\Hf ,\Hb ]
      &=& \left[ \ary{cc} {
              A & B \\ -B & -A
              }\right]\ .
             \label{commutator matrix in A and B}
\eeqa
The eigenvalues of the last matrix are $\pm\sqrt{A^2-B^2}$.   Note that $A$ is imaginary
and $B$ is real, and so the eigenvalues are imaginary. We now determine empirical values
for the variables $M_{i,j}$ in \eq{commutator matrix}.

The phenomenological model given in Lee and Wolfenstein \cite{Lee} describes the
evolution of the amplitudes $a(t)$ and $b(t)$ for the single-particle kaon field to be in
the states \raisebox{0pt}[0pt]{$\ket{K^0}$} and \raisebox{0pt}[0pt]{$\ket{\bK{}^0}$},
respectively, by the differential equation
\beq
    \label{diff eq Lee&Wolf}
     \frac{d\psi(t)}{dt}=-\frac{1}{\hbar}(\Gamma + i M)\psi(t)
\eeq
where $\psi=\left(^{a}_{b}\right)$.  The matrix on the right-hand side is defined by
\beqa
     \Gamma + i M &=& D+i(E_1\sigma_1+E_2\sigma_2+E_3\sigma_3)
     \label{Gamma + i M}
\eeqa
where the matrices $\Gamma$ and $M$ are Hermitian and
\beqa
     \sigma_1=\left[\ary{cc} {0 & 1\\ 1 & 0}\right]\ ,\qquad
     \sigma_2=\left[\ary{cc} {0 & -i\\ i & 0}\right]\ ,\qquad
     \sigma_3=\left[\ary{cc} {1 &0\\0&-1}\right]\ .
     \label{sigmas}
\eeqa
The analysis below shows that the matrix $M$ in \eq{diff eq Lee&Wolf}
represents a CP violating Hamiltonian. For our purposes, it provides a
sufficient estimate of the T violating Hamiltonian for a neutral kaon.
Accordingly, we use the empirical values of the elements of $M$ as estimates
of the corresponding matrix elements $M_{i,j}$ of the Hamiltonian $\Hf$ that
appears in \eq{commutator matrix}. We should also mention that we could use
$M$ as an estimate of the matrix elements of $\Hb$ instead, however the only
difference this would make is that the matrix elements of the commutator
$[\Hf ,\Hb ]$ would be multiplied by $-1$, and this has no physical
consequences because the two eigenvalues are the negatives of each other.

If the CPT theorem holds, which we assume to be the case, then $E_1=E\cos\phi$,
$E_2=E\sin\phi$ and $E_3=0$. Here $\phi$ is a complex parameter \cite{Lee} which is
related to the CP violation parameter $\epsilon$ by
$\epsilon=(1-e^{i\phi})/(1+e^{i\phi})$, or equivalently,
\beqa
     e^{i\phi}&=&\frac{1-\epsilon}{1+\epsilon}\ .
     \label{exp(-i phi)}
\eeqa
The values of $D$ and $E$ comprise the decay rates $\gamma_i$ and masses $m_i$ of the
short and long-lived components, $\ket{K_1^0}$ and $\ket{K_2^0}$ respectively, of
$\ket{K^0}$ as follows
\beqa
    D&=&\sfc{1}{4}(\gamma_1+\gamma_2)+\sfc{1}{2}i(m_1+m_2)
    \label{D expression}\\
    iE&=&\sfc{1}{4}(\gamma_1-\gamma_2)+\sfc{1}{2}i(m_1-m_2)\ .
    \label{iE expression}
\eeqa
Here $\gamma_i$ and $m_i$ are the energies $\gamma_i=\hbar\gamma'_i$ and $m_i=m'_i c^2$,
respectively, where $\gamma'_i$ and $m'_i$ are the actual decay rates and masses,
respectively. Also note \cite{Yao} that $|\epsilon|\approx 2.3\times 10^{-3}$ and
$\arg(\epsilon)\approx 45^\circ$ and so from \eq{exp(-i phi)} we find
$e^{i\phi}=1-2\epsilon+{\cal O}(\epsilon^2)$.  We can therefore make the following
approximations
\beqan
     e^{i\phi}&=&1-\sqrt{2}|\epsilon|(1+i)+{\cal
     O}(\epsilon^2)\non\ ,\\
     e^{i\phi^*}&=&1/(e^{i\phi})^*=1+\sqrt{2}|\epsilon|(1-i)+{\cal
     O}(\epsilon^2)\non\ ,\\
     e^{-i\phi}&=&1/(e^{i\phi})=1+\sqrt{2}|\epsilon|(1+i)+{\cal
     O}(\epsilon^2)\non\ ,\\
     e^{-i\phi^*}&=&\ (e^{i\phi})^*\;=1-\sqrt{2}|\epsilon|(1-i)+{\cal
     O}(\epsilon^2)\ .
\eeqan
This allows us to write down the following useful results
\beqan
     \sfc{1}{2}(e^{i\phi}+e^{i\phi^*})&=&1-i\sqrt{2}|\epsilon|+{\cal
     O}(\epsilon^2)\ ,\\
     \sfc{1}{2}(e^{-i\phi}+e^{-i\phi^*})&=&1+i\sqrt{2}|\epsilon|+{\cal
     O}(\epsilon^2)\ ,\\
     \sfc{1}{2}(e^{i\phi}-e^{i\phi^*})&=&-\sqrt{2}|\epsilon|+{\cal
     O}(\epsilon^2)\ ,\\
     \sfc{1}{2}(e^{-i\phi}-e^{-i\phi^*})&=&\sqrt{2}|\epsilon|+{\cal
     O}(\epsilon^2)\ .
\eeqan
From Eqs.~(\ref{Gamma + i M}), (\ref{sigmas}), (\ref{D expression}) and (\ref{iE
expression}) we find
\beqa
     \Gamma + i M &=& D+i(E\cos\phi\;\sigma_1+E\sin\phi\;\sigma_2)\non\\
     &=&\left[\ary{cc} {
              \sfc{1}{4}(\gamma_1+\gamma_2)+\sfc{1}{2}i(m_1+m_2) & \Big[\sfc{1}{4}(\gamma_1-\gamma_2)+\sfc{1}{2}i(m_1-m_2)\Big]e^{-i\phi}\\
              \Big[\sfc{1}{4}(\gamma_1-\gamma_2)+\sfc{1}{2}i(m_1-m_2)\Big]e^{i\phi} &\sfc{1}{4}(\gamma_1+\gamma_2)+\sfc{1}{2}i(m_1+m_2)
     }\right]\ .\non \\
     \label{Gamma +iM matrix}
\eeqa
The Hermitian transpose of this equation is
\beqa
     \Gamma - i M
     &=&\left[\ary{cc} {
              \sfc{1}{4}(\gamma_1+\gamma_2)-\sfc{1}{2}i(m_1+m_2) & \Big[\sfc{1}{4}(\gamma_1-\gamma_2)-\sfc{1}{2}i(m_1-m_2)\Big]e^{-i\phi^*}\\
               \Big[\sfc{1}{4}(\gamma_1-\gamma_2)-\sfc{1}{2}i(m_1-m_2)\Big]e^{i\phi^*}  &  \sfc{1}{4}(\gamma_1+\gamma_2)-\sfc{1}{2}i(m_1+m_2)
     }\right]\ .\non\\
     \label{Gamma -iM matrix}
\eeqa
By subtracting \eq{Gamma -iM matrix} from \eq{Gamma +iM matrix} and dividing the result
by $2i$ we get
\beqa
     M&=&\left[\ary{cc} {
              \sfc{1}{2}(m_1+m_2) & -i\sfc{1}{4}(\gamma_1-\gamma_2)\sqrt{2}|\epsilon|+\sfc{1}{2}(m_1-m_2)(1+i\sqrt{2}|\epsilon|)\\
               i\sfc{1}{4}(\gamma_1-\gamma_2)\sqrt{2}|\epsilon|+\sfc{1}{2}(m_1-m_2)(1-i\sqrt{2}|\epsilon|)  &  \sfc{1}{2}(m_1+m_2)
     }\right]\non\\
     &&\ +{\cal O}(\epsilon^2)\ .
\eeqa
This shows that $M_{11}=M_{22}$ and so $B=0$ in Eqs.~(\ref{commutator matrix}) and
(\ref{commutator matrix in A and B}). The eigenvalues of the commutator $[\Hf ,\Hb ]$ are
therefore simply $\pm A$ where
\beqa
    A&=&M_{12}^2-M_{21}^2  \non\\
    &=&\Big[-i\sfc{1}{4}(\gamma_1-\gamma_2)\sqrt{2}|\epsilon|+\sfc{1}{2}(m_1-m_2)(1+i\sqrt{2}|\epsilon|)\Big]^2\non\\
    &&\hs{5}-\Big[i\sfc{1}{4}(\gamma_1-\gamma_2)\sqrt{2}|\epsilon|+\sfc{1}{2}(m_1-m_2)(1-i\sqrt{2}|\epsilon|)\Big]^2
        +{\cal O}(\epsilon^2)\non\\
    &=& -i\sfc{1}{4}(\gamma_1-\gamma_2)\sqrt{2}|\epsilon|\sfc{1}{2}(m_1-m_2)\Big[2(1+i\sqrt{2}|\epsilon|)+2(1-i\sqrt{2}|\epsilon|)\Big]\non\\
    &&\hs{5}  +\sfc{1}{4}(m_1-m_2)^2\Big[(1+i\sqrt{2}|\epsilon|)^2-(1-i\sqrt{2}|\epsilon|)^2\Big]
                +{\cal O}(\epsilon^2)\non\\
    &=& i\Big[-(\gamma_1-\gamma_2)\sqrt{2}|\epsilon|\sfc{1}{2}(m_1-m_2)
      +(m_1-m_2)^2\sqrt{2}|\epsilon|\Big]  +{\cal
      O}(\epsilon^2)\non\\
    &=& i\sqrt{2}|\epsilon|\Big[(\gamma_1-\gamma_2)\sfc{1}{2}(m_2-m_1)
      +(m_1-m_2)^2\Big]  +{\cal O}(\epsilon^2)\non\\
    &=& i\sqrt{2}|\epsilon|\Big[\sfc{1}{2}\Delta\gamma\Delta m+(\Delta m)^2\Big] +{\cal O}(\epsilon^2)
    \label{A = delta m + delta gamma}
\eeqa
where $\Delta m =(m_2-m_1)$ and $\Delta \gamma=(\gamma_1-\gamma_2)$.

Finally, the empirical values \cite{Yao}
\beqan
  \Delta m &\approx0.56\times 10^{10}\hbar s^{-1}\ ,\\
  \Delta \gamma&\approx 1.1\times10^{10}\hbar s^{-1}\ ,
\eeqan
show that the eigenvalue $A$ in \eq{A = delta m + delta gamma} can be approximated as
\beqa
    A&\approx i10^{17}\hbar^2 s^{-2}\ .
\eeqa
This is the value used in the paper for $\lambda ^{(1)}$ where $\lambda ^{(1)}=\pm iA$.
Note that in the main text, we use units in which $\hbar=1$. This is equivalent to
replacing the Hamiltonians $H_{F}$ and $H_{B}$ in this section with $H_{F}/\hbar$ and
$H_{B}/\hbar$, respectively.  In that case the commutator in \eq{commutator matrix in A
and B} has the form $[\Hf ,\Hb ]/\hbar^2$ which has a dimension of (time)$^2$.  Hence
$\lambda ^{(1)}\approx \pm 10^{17}\,s^{-2}$.

\end{document}